\documentclass[aps,pra,twocolumn,showpacs,groupedaddress]{revtex4}
\usepackage{graphicx}

\begin{document}

\newcommand{\Singlet}{X^1\Sigma^+}
\newcommand{\Triplet}{a^3\Sigma^+}


\title{Calculation of the interspecies s-wave scattering length in
an ultracold Na-Rb vapor}

\author{S. B. Weiss, M. Bhattacharya, and  N. P. Bigelow}
\address{Department of Physics and Astronomy, The Laboratory for Laser Energetics,
and The Institute of Optics\\ The University of Rochester,
Rochester, NY 14627}

\date{\today}

\begin{abstract}
We report the calculation of the interspecies scattering length
for the sodium-rubidium (Na-Rb) system.  We present improved
hybrid potentials for the singlet $\Singlet$ and triplet
$\Triplet$ ground states of the NaRb molecule, and calculate the
singlet and triplet scattering lengths $a_{s}$ and $a_{t}$ for the
isotopomers $^{23}$Na$^{87}$Rb and $^{23}$Na$^{85}$Rb. Using these
values, we assess the prospects for producing a stable two-species
Bose-Einstein condensate in the Na-Rb system.
\end{abstract}
 \maketitle

\section{Introduction}

The $s$-wave scattering length $a$ plays a central role in the
description of atom-atom collisions at ultralow temperatures ($T$
$\ll$ 1 mK). In this regime, the cross section for elastic
collisions, $\sigma_{el} \sim \pi a^2$, and the cross section for
inelastic spin-exchange collisions, $\sigma_{ex} \sim \pi
({a_t}-{a_s})^2$, are both expressed in terms of $a$
\cite{Weiner1999Julienne-Collisions}. The scattering length is
also a critically important parameter in the physics of
Bose-Einstein condensates (BECs).  For a bosonic atomic species
{\em i}, a BEC is stable only if $a_i > 0$. In addition, efficient
evaporative cooling demands that $\sigma_{el} \gg \sigma_{ex}$
\cite{EvapCool}.  {\em A priori} calculations of $a$ are thus of
fundamental interest, and the quest for BEC in alkali-metal atoms
has spurred on efforts to calculate the scattering length in many
atomic species.

A majority of the work on scattering lengths has concentrated on
interactions between like alkali-metal atoms.  However, the recent
production of Bose-Fermi mixtures in $^{6}$Li-$^{23}$Na
\cite{Hadzibabic2002Ketterle-LiNa}, $^{6}$Li-$^{7}$Li
\cite{Truscott2001Hulet-6Li7Li,Schreck2001Salomon-6Li7Li}, and
$^{40}$K-$^{87}$Rb \cite{Roati2002Inguscio} and dual-species BEC
in $^{41}$K-$^{87}$Rb \cite{Modugno2002Inguscio-TBEC1} has renewed
interest in binary mixtures of ultracold gases.  In these systems,
the interspecies scattering length $a_{12}$ is the basic quantity
parametrizing the interactions between component atomic species.
$a_{12}$ determines the efficiency of sympathetic cooling
$\emph{en route}$ to the formation of quantum degenerate mixtures,
and in the case of two-species BECs (TBECs), $a_{12}$ determines
the stability and miscibility of these mixtures \cite{TBEC-NaRb}.
However, relatively few calculations of interspecies scattering
lengths exist \cite{Venturi2001Cote-NaK,InterSpeciesSL}. This is
due, in part, to the incomplete characterization of diatomic
interaction potentials for many pairs of alkali-metal atoms.

In this paper we calculate the scattering lengths for the Na-Rb
system.  We construct the NaRb potential from a combination of
spectroscopic data and precise long-range interaction parameters,
and use a simple method for calculating the singlet and triplet
scattering lengths for the isotopomers $^{23}$Na$^{87}$Rb and
$^{23}$Na$^{85}$Rb.  The Na-Rb system is interesting in part
because its an obvious candidate for TBEC. Both of the component
species have been condensed and the condensates have been studied
in detail \cite{BECreviews}. The Na-Rb TBEC has been treated
theoretically, and its properties are sensitive to the value of
the interspecies scattering length $a_{Na-Rb}$ \cite{TBEC-NaRb}.

This paper is divided into four sections.  In
Sec.~\ref{Section:SL}, we discuss methods for calculating
$a_{Na-Rb}$, and address the source of errors involved in these
calculations. In Sec.~\ref{Section:Potentials}, we introduce two
hybrid potentials for the NaRb molecule, and contrast our
potentials with previous results. In Sec.~\ref{Section:Results},
we present our calculations for $a_{Na-Rb}$, and discuss the
feasibility of producing a TBEC in the Na-Rb system.

\section{Scattering Length\label{Section:SL}}

If the potential $V(r)$ is known for all $r$, then the scattering
length $a$ can be calculated ~\cite{Joachain}.  The procedure is
to numerically integrate the radial Schrodinger equation for low
collision energies to large values of $r$. At large $r$ the
numerical wave function $u(r)$ is ``matched" to an asymptotic form
$\chi(r)$ by requiring that $\frac{\partial}{\partial r}\ln u(r) =
\frac{\partial}{\partial r}\ln \chi(r)$ at the match point
$r=r_m$.  The asymptotic wave function can be written as
\begin{equation} \label{eq:PsiLR} \chi_{l=0}(r)
\sim{1\over{k}}[\sin(kr)+\tan(\delta_{o})\cos(kr)],
\end{equation}
where $k$ is the atomic wave vector and $\delta_{o}$ is the
$s$-wave phase shift. In the limit that $k \rightarrow 0$, the
wave function asymptotically approaches a straight line as $r
\rightarrow \infty$.  The scattering length is given by the $r$
intercept of this line \cite{SLfootnote1}, and can be formally
defined in terms of the $s$-wave phase shift as
\begin{equation}
\label{eq:sldef}
a=-\lim_{k\rightarrow0}\frac{\tan(\delta_{o})}{k}.
\end{equation}
For low collision energies, $r_m$ should be large to ensure that
the numerical wave function attains its asymptotic behavior.  The
total integration time can be reduced by calculating the
corrections to the asymptotic wave function at smaller values of
the match point $r=r_m$.  For example, Marinescu has shown
\cite{Marinescu1994-SL} that the wave function for large $r$ may
be written as $\chi(r) = \alpha \epsilon_{\alpha}(r) + \beta
\epsilon_{\beta}(r)$, where the functions
$\epsilon_{\alpha,\beta}(r)$ are solutions to the differential
equations $\epsilon_{\alpha,\beta}''(r)= [2 \mu V(r)
/\hbar^2]\epsilon_{\alpha,\beta}(r)$ subject to the boundary
conditions $\epsilon_{\alpha}(r) \rightarrow r$ and
$\epsilon_{\beta}(r) \rightarrow 1$ as $r \rightarrow \infty$.
These equations cannot be solved exactly. However, if an analytic
expression for the long-range potential is known, then the
functions $\epsilon_{\alpha,\beta}(r)$ may be estimated to
arbitrary precision using a method of successive approximations.
The scattering length is given by $a=-\beta/\alpha$, and can be
found by applying the usual continuity condition at $r=r_m$.

The uncertainty in the calculated value of $a$ depends upon the
reliability of the potential $V(r)$.  Gribakin and Flambaum
\cite{Gribakin1993Flambaum-SL} have shown that for diatomic
potentials which fall off at long range as $-C_{6}/r^6$, the
scattering length is given by
\begin{equation}
\label{eq:gfsl} a=a_{o}[1-\tan\{\Phi(E=0)-\pi/8\}],
\end{equation}
where $a_{o}$ is a``mean scattering length", and $\Phi(E)$ is the
semiclassical phase, defined as
\begin{equation}
\label{eq:Phase} \Phi(E)=\int_{r_{inner}}^{r_{outer}}
\sqrt{\frac{2\mu(E- V(r))}{\hbar^2}}\,dr ,
\end{equation}
where $r_{inner}$ ($r_{outer}$) is the inner (outer) classical
turning point of the potential at energy $E$ and $\mu$ is the
reduced mass of the colliding atoms.  As can be seen from Eq.
(\ref{eq:gfsl}), the scattering length is infinite if
$\Phi(E=0)=(n - 3/8) \pi$, where ($n$=1,2,3,...).  This situation
occurs if the $n{\rm{th}}$ vibrational state of the potential
$V(r)$ is barely bound at $E=0$.  In general, $V(r)$ will not
admit a barely bound state. However, variations of the potential
within its estimated uncertainties will shift the energies of its
bound states, and states lying closest to dissociation experience
the largest shifts.  A bound state may even be introduced or
removed from the well, depending on the size of the potential
shift and the proximity of a bound or virtual level to the
dissociation energy. As states are added or removed from the well,
the scattering length passes through $\pm \infty$. Therefore, if
the interatomic potential is not known well enough to predict
whether or not a barely bound state exists, then $a$ cannot be
specified within finite bounds.

Because of the extreme sensitivity of $a$ to the binding energy of
the highest vibrational state of the interatomic potential $V(r)$,
the most precise calculations of $a$ in alkali-metal atoms
typically rely on the spectroscopy of bound states near
dissociation. Two-color photoassociation or Raman spectroscopy is
used to resolve these lines to high precision
\cite{Tsai1997Heinzen-2ColorPAS}.  In the absence of
near-dissociation spectroscopy, the scattering length may still be
calculated, but the accuracy and precision of such a calculation
is limited by the accuracy and precision of the interatomic
potential $V(r)$.  In the case of alkali-metal dimers,
spectroscopy is sparse and near-dissociation spectroscopy is
nonexistent. However, the potential $V(r)$ may still be
``assembled" from RKR (Rydberg-Klein-Rees) \cite{RKRreferences}
data and well-known analytic expressions for short- and long-range
potentials to create a ``hybrid" potential valid for all $r$.  In
assigning error bars to $a$, care must be taken to ensure that
variations of the hybrid potential within its estimated
uncertainties do not introduce or remove bound states from the
well.  For example, uncertainties in the $C_{6}$ coefficient of
NaK allowed for additional bound states in its hybrid potential,
and frustrated attempts to determine the scattering length for
some isotopomers \cite{Venturi2001Cote-NaK}. Fortunately, this is
not the case for NaRb, as shown below.

\section{N$\lowercase{\rm{\bf{a}}}$R$\lowercase{\rm{\bf{b}}}$ Potentials\label{Section:Potentials}}

Compared to many alkali-metal dimers, the ground states of the
NaRb molecule are relatively well-known.  Rovibrational states to
within 5$\%$ of dissociation have been observed in both the
triplet $\Triplet$ \cite{Wang1991Kato-NaRb} and singlet $\Singlet$
potential wells
\cite{Kasahara1996Kato-NaRb,Docenko2002Stolyarov-NaRb}. The
rotationless interatomic potentials $V_s(r)$ and $V_t(r)$,
corresponding to the $\Singlet$ and $\Triplet$ states,
respectively, have been determined through RKR analysis, and a
direct fit to the singlet spectrum
\cite{Docenko2002Stolyarov-NaRb} using a modified Lennard-Jones
(MLJ) \cite{Hajigeorgiou2002LeRoy-MLJ} parametrization has also
been performed. {\em Ab initio} ground state potential curves for
NaRb have been calculated \cite{Korek2000Frecon-NaRb}.  However,
these curves are not very accurate, so we do not use them in
constructing our interatomic potentials.

At large $r$, the NaRb interatomic potential is accurately
represented by a sum of two independent contributions, the
exchange and dispersion energies.  The dispersion energy is given
by a well-known expansion in powers of $r^{-1}$:
\begin{equation}
\label{eq:dispersion}
V_{disp}(r)=-\biggl(\frac{C_6}{r^6}+\frac{C_8}{r^8}+\frac{C_{10}}{r^{10}}\biggr).
\end{equation}
The coefficients $C_{n}$ may be calculated from a knowledge of
atomic polarizabilities \cite{Stone}.  The exchange interaction is
calculated using the surface integral method of Smirnov and
Chibisov \cite{Smirnov1965Chibisov-Vex}, which yields
\begin{equation}
\label{eq:Vex} V_{ex}(r)=\pm \frac{1}{2} J(A,B,\alpha,\beta,r)\,\,
r^{\frac{2}{\alpha} + \frac{2}{\beta} - \frac{1}{(\alpha + \beta)}
-1} e^{-(\alpha + \beta)r},
\end{equation}
where $\alpha^{2}/2$ and $\beta^{2}/2$ are the ionization energies
(in atomic units) of each atom, and $r$ is assumed to be in units
of Bohr radii. The function $J(A,B,\alpha,\beta,r)$ can be
expanded in a power series $\sum_{n}(J_n r^n (\alpha-\beta)^n)/n!$
whose coefficients $J_n$ are expressed as integrals that must be
solved numerically.  The complete long-range potential is then
given by
\begin{equation}
\label{eq:longrange} V_{LR}(r) = V_{disp}(r) \pm V_{ex}(r).
\end{equation}

The exchange energy is positive (negative) for the triplet
(singlet) state.  As $r \rightarrow \infty$, the long-range
interaction potential is dominated by the well-known van der Waals
potential $-C_6/r^6$. The exchange interaction is expected to
become important inside the LeRoy radius $R_{LeRoy}$
\cite{LeRoyReview}, beyond which the potential is well
approximated by the dispersion energy alone. For NaRb, $R_{LeRoy}
\sim$ 11 $\rm\AA$.

The NaRb molecular potentials can be modelled by smoothly joining
RKR data to the long-range interaction potentials. Zemke and
Stwalley (Z-S) have constructed such hybrid potentials for the
$\Triplet$ and $\Singlet$ states of the NaRb molecule
\cite{Zemke2001Stwalley-NaRb}.  More complete spectra of the NaRb
singlet state \cite{Docenko2002Stolyarov-NaRb} and a more precise
estimate of the $C_6$ coefficient for NaRb
\cite{Derevianko2001Dalgarno-HeteroC6} have since become
available, allowing us to construct new hybrid potentials for both
the NaRb $X^1\Sigma^+$ and $a^3\Sigma^+$ states.  For $r < 11\
\rm\AA$, we use the recent MLJ potential to model the
$X^1\Sigma^+$ state. Our potential for the $\Triplet$ state is
identical to the Z-S triplet potential for $r< 13.5788\ \rm\AA$.
Our long-range potential differs from that used by Z-S in a number
of ways.  We use the $C_8$ and $C_{10}$ dispersion coefficients
recommended by Marinescu and Sadeghpour
\cite{Marinescu1999Sadeghpour-LR}, but choose for $C_6$ the highly
precise value calculated by Derevianko \emph{et al.}
\cite{Derevianko2001Dalgarno-HeteroC6}.  For the exchange energy,
we used the heteronuclear expression of Smirnov and Chibisov given
by Eq. (\ref{eq:Vex}). Both the singlet MLJ and triplet RKR
potential curves are joined smoothly to our long-range potential
as given by Eq. (\ref{eq:longrange}). Our complete long-range
potential is given in Table~\ref{tab:parameterLR}.
\begin{table}
\caption{\label{tab:parameterLR} Our chosen values for the
parameters of the NaRb long-range potential $V_{LR}(r)$.}
\begin{ruledtabular}
\begin{tabular}{cc}
Parameter& Value\\ \hline $C_6$ \footnotemark[1] & 1.293$\times
10^7$
\\ $C_8$ \footnotemark[2] & 3.4839$\times 10^8$ \\ $C_{10}$ \footnotemark[2] &
1.1552$\times 10^{10}$
\\ $\alpha_{Na}$ \footnotemark[3] & 0.61459 \\ $\beta_{Rb}$ \footnotemark[3]
& 0.55409 \\ $A_{Na}$ \footnotemark[4] & 0.76752 \\ $B_{Rb}$
\footnotemark[4] & 0.56945
\\ $J_0$ \footnotemark[5] & 1.4197$\times 10^{-2}$ \\ $J_1$ \footnotemark[5] & 6.0963$\times 10^{-4}$ \\
$J_2$ \footnotemark[5] & 1.9537$\times 10^{-3}$ \\
\end{tabular}
\end{ruledtabular}
\footnotetext[1]{$C_6$ given in units of $\rm{cm}^{-1}\
\rm{\AA}^6$. See Ref. \cite{Derevianko2001Dalgarno-HeteroC6}.}
\footnotetext[2]{$C_8$ and $C_{10}$ given in units of
$\rm{cm}^{-1}\ \rm{\AA}^8$ and $\rm{cm}^{-1}\ \rm{\AA}^{10}$,
respectively.  See Ref. \cite{Marinescu1999Sadeghpour-LR}.}
\footnotetext[3]{The quantities are expressed in atomic units. See
Ref. \cite{NISTwebsite}.} \footnotetext[4]{The constants $A$ and
$B$ are related to the size of the wavefunction of each atom in
the region of interaction. See Ref.
\cite{Marinescu1996Dalgarno-LRHF}.} \footnotetext[5]{We found that
the exchange energy was adequately represented in our region of
interest by the first three terms of the expansion
$J(A,B,\alpha,\beta,r) = \sum_{n}(J_n r^n (\alpha-\beta)^n)/n!$.
Here we use atomic units. See Ref.
\cite{Smirnov1965Chibisov-Vex}.}
\end{table}

\section{Results and Discussion\label{Section:Results}}

\subsection{Na-Rb scattering lengths}

To calculate the singlet or triplet scattering lengths, we choose
the hybrid potential $V_s(r)$ or $V_t(r)$ and integrate the radial
Schrodinger equation at $E=0$ from $r_{inner}$ to the match point
$r_{m}$ using the Numerov algorithm ~\cite{Numerov}.  The reduced
mass $\mu$ is given by $M_{Na}M_{Rb}/(M_{Na}+M_{Rb})$, where $M$
labels the atomic mass of either $^{23}$Na, $^{85}$Rb, or
$^{87}$Rb.  We expect the isotopic correction to the internuclear
potential $V(r)$ to be negligible
\cite{Docenko2002Stolyarov-NaRb}.  Following Ref.
\cite{Marinescu1994-SL}, the scattering length is given by
\begin{equation}
a=\frac{u \epsilon_{\alpha}' - u' \epsilon_{\alpha}}{u
\epsilon_{\beta}' - u' \epsilon_{\beta}}~\Bigr|_{r=r_m},
\end{equation}
where $u(r)$ is the numerically integrated wave function, and the
functions $\epsilon_{\alpha,\beta}(r)$ are determined from the
long-range potential, as discussed in Sec.~\ref{Section:SL}. The
primes denote derivatives with respect to $r$.  We found that a
fourth-order approximation to $\epsilon_{\alpha}(r)$ and
$\epsilon_{\beta}(r)$ guaranteed convergence to a reliable value
of $a$ at a match point $r_m = 100\ \rm{\AA}$. Our results are
summarized in Table~\ref{tab:SL}.
\begin{table}
\caption{\label{tab:SL} Scattering lengths found from our hybrid
potentials for $^{23}$Na$^{85}$Rb and $^{23}$Na$^{87}$Rb, in units
of Angstroms. }
\begin{ruledtabular}
\begin{tabular}{ccc}
Isotopomer&$a_S$&$a_T$\\ \hline $^{23}$Na$^{85}$Rb &
$167_{-30}^{+50}$ & $59_{-9}^{+12}$\\ $^{23}$Na$^{87}$Rb &
$55_{-3}^{+3}$ & $51_{-6}^{+9}$\\
\end{tabular}
\end{ruledtabular}
\end{table}

Because the scattering length is very sensitive to the details of
the interatomic potential $V(r)$, it is important to ensure that
our calculated values of $a$ are stable with respect to changes in
$V_{s,t}(r)$ within their known experimental or theoretical
uncertainties. These include uncertainties in the value of the
dissociation energy $D_e$, the inner and outer turning points of
the RKR potentials, the binding energy of the observed vibrational
states, and the coefficients of the long-range parameters. We
estimate our errors by calculating the change in the semiclassical
phase $\Phi(E=0)$ due to the error in each parameter of our
potential. These ``phase errors" $\Delta \phi$ are then summed in
quadrature to give a total phase error $\Delta \phi_{total}$.  We
convert this value into a scattering length error using Eq.
(\ref{eq:gfsl}).

For the triplet state, the error in $a$ was estimated with respect
to the uncertainties in $D_e$, $C_6$, and the RKR turning points.
In the case of the singlet state, the MLJ parametrization allowed
us to simultaneously vary all parameters in a statistically
meaningful way. We treated the MLJ parameters as random variables
with a well-defined mean and standard deviation. The phase
$\Phi(E=0)$ was then calculated for 100 ``random" potentials. The
phase error was determined by examining the distribution of
phases. This phase error was used to calculate the error in $a_s$.

As shown in Table~\ref{tab:SL}, our error bars are small.  This
is, in some ways, a fortuitous result. Had a bound or virtual
state been closer to dissociation, variations in the potential may
have caused the phase to pass through a region where $a
\rightarrow \infty$.  Because the scattering lengths are
relatively small, they are more stable with respect to changes in
the corresponding potential.  In addition, our hybrid potentials
are reasonably well constrained.  This is due, in part, to the
observation of bound states relatively close to dissociation in
the NaRb triplet well, which enabled Z-S to reduce the uncertainty
in the dissociation energy to $\Delta(D_{e})=\pm
0.1~\rm{cm}^{-1}$, and the very precise calculation of the $C_{6}$
coefficient by Derevianko \cite{Derevianko2001Dalgarno-HeteroC6}.

\begin{figure}
\includegraphics[width=8.5cm]{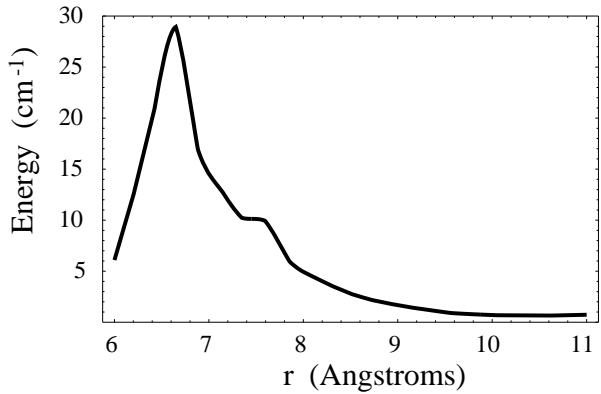}
\caption{\label{fig:MLJ-Zsdiff} A plot of the difference between
the Z-S and MLJ $\Singlet$ potentials.}
\end{figure}

One interesting consequence of the new singlet potential is the
appearance of an additional bound state.  We found that our
complete potential for the $\Singlet$ state supported 83 bound
states, whereas the corresponding Z-S potential only supported 82
bound states. This can be understood by examining
Fig.~\ref{fig:MLJ-Zsdiff}, which shows the energy difference
between the Z-S singlet and MLJ potentials for $6\ \rm\AA < r <
11\ \rm\AA$.
Because the MLJ potential is everywhere deeper, the wave vector
$k(E=0)$ will be larger for all $r$.  In semiclassical terms, the
wave function will build up more phase $\Phi$ in this potential.
In our case, $\Phi_{MLJ}-\Phi_{Z-S} \approx \pi$, so that another
bound state appears in the MLJ molecular well. The energy
differences between the two curves is due primarily to errors in
the extrapolation procedure used by Z-S to connect the short- and
long-range components of their hybrid singlet state potential.

\subsection{Two-species Na-Rb BEC}

Having calculated the Na-Rb scattering lengths, we now consider
the properties of a mixture of Na and Rb condensates.  The
interactions in a Na-Rb TBEC are parametrized by three scattering
lengths: $a_{Na}$, $a_{Rb}$, and $a_{Na-Rb}$. The values of
$a_{Rb}$ and $a_{Na}$ are known to high precision, and are given
in Table~\ref{tab:Na&RbSL}.
\begin{table}
\caption{\label{tab:Na&RbSL} Recently determined scattering
lengths for $^{23}$Na, $^{85}$Rb, and $^{87}$Rb, in units of Bohr
radii.}
\begin{ruledtabular}
\begin{tabular}{lcr}
Species&$a_s$&$a_t$\\ \hline $^{23}$Na \footnotemark[1] & $19.1
\pm 2.1$ & $65.3 \pm 0.9$\\ $^{85}$Rb \footnotemark[2] &
$2795_{-290}^{+420}$ & $-388 \pm 3$\\ $^{87}$Rb \footnotemark[2] &
$90.4 \pm 0.2$ & $98.98 \pm 0.04$\\
\end{tabular}
\end{ruledtabular}
\footnotetext[1]{See Ref. \cite{vanAbeelen1999Verhaar-NaSL}.}
\footnotetext[2]{See Ref. \cite{vanKempen2002Verhaar-Rb}.}
\end{table}
We consider a Na-Rb mixture in which the Na atoms are in the state
$|F_{Na},m_{Na}\rangle$, and the Rb atoms are in the state
$|F_{Rb},m_{Rb}\rangle$, where $F$ is the total angular momentum
of the atom and $m$ is its projection on the quantization axis. We
represent the two-atom state by the ket
$|F_{Na},m_{Na};F_{Rb},m_{Rb}\rangle$.  To calculate the
scattering lengths we use the low-energy elastic approximation
\cite{ElasticApprox,Cote1998Stwalley-ElasticApprox}, which assumes
that elastic collisions dominate the total cross section for
collisions.  This approximation gives for the scattering lengths
\begin{equation}
\label{eq:elasticSL}
    a=a_{s}P_{s}+a_{t}P_{t},
    \label{1}
\end{equation}
where $P_s$ and $P_t$ are the probabilities of the atoms being in
a singlet or triplet state, respectively. To calculate the
probabilities $P_{s}$ and $P_{t}$, we project the state
$|F_{Na},m_{Na};F_{Rb},m_{Rb} \rangle$ onto the states
$|S,m_{S};I,m_{I}\rangle$, where $S$ and $I$ refer to the total
electronic and nuclear spin of the two-atom system, respectively,
while $m_{S,I}$ are their projections onto the quantization axis.
This basis is useful for characterizing the system at smaller
internuclear distances where the exchange energy dominates.  In
this region, $F_{Na}$ and $F_{Rb}$ are no longer ``good" quantum
numbers, and the singlet and triplet states are labeled by
$S=0,1$, respectively.

To calculate the projections
$C_{S,m_S,I,m_I}^{F_{Na},m_{Na},F_{Rb},m_{Rb}} = \langle S, m_S ;
I, m_I | F_{Na},m_{Na}; F_{Rb}, m_{Rb} \rangle $, we perform the
angular momentum recoupling of the four quantum numbers $S_{Na}$,
$I_{Na}$, $S_{Rb}$ and $I_{Rb}$ by making use of the Wigner 9-j
symbols and standard Clebsch-Gordan algebra.  We calculated the
complete recoupling matrices $U_{SI-FF}$ for both
$^{23}$Na$^{85}$Rb and $^{23}$Na$^{87}$Rb, and extracted the
probabilities $P_s$ and $P_t$ for all input channels
$|F_{Na},m_{Na};F_{Rb},m_{Rb}\rangle$.  The scattering length for
an arbitrary input channel is then given by Eq.
(\ref{eq:elasticSL}).

Knowledge of $a$ enables us to calculate the cross section for
elastic collisions, $\sigma_{el}=4 \pi a^2$. Elastic collisions
mediate the rethermalization of atoms during evaporative cooling
and sympathetic cooling. Furthermore, knowledge of $a_s$ and $a_t$
allows us to characterize the inelastic losses in the system. The
dominant two-body mechanism for the loss of atoms from a trap is
spin-exchange collisions.  In this type of collision, the internal
spin states of one or both of the atoms changes.  In the case of
magnetically trapped samples, such collisions can cause atoms to
be ejected from the mixture by sending them into nontrappable spin
states. More generally, the atoms may be ejected if the spin
reorientation energy is converted to kinetic energies greater than
the depth of the (magnetic or optical) potential confining the
mixture.  In the elastic approximation we can write the cross
section for such inelastic processes as
\cite{Cote1998Stwalley-ElasticApprox}
\begin{equation}
    \sigma_{ex} = M_{if} \pi(a_{t}-a_{s})^{2},
\end{equation}
where $M_{if}$ is a factor that depends on the asymptotic
hyperfine states involved in the collision.  Letting primes denote
the asymptotic output channel, we have
\begin{equation}
    M_{if}=\Biggl [ \sum_{m_S , I , m_I} (C_{S=0}C'_{S=0} -
    C_{S=1}C'_{S=1}) \Biggr ]^2,
\end{equation}
where $C$ is the projection coefficient defined above, and the
indices have been suppressed.

The achievement of a miscible two-component BEC places a number of
constraints on the three relevant scattering lengths. Efficient
sympathetic cooling requires a large magnitude of $a_{Na-Rb}$.
Collisional stability against spin-exchange collisions requires
small values of $\sigma_{ex}$, which implies that the difference
between $a_s$ and $a_t$ be small.  Dynamical stability of the
individual BECs requires $a_{Na}>0$ and $a_{Rb}>0$.  In the
Thomas-Fermi approximation, the criteria for stability implies the
existence of a critical value of $|a_{Na-Rb}| = a_c $ above which
the two-species condensate cannot coexist.  The criteria is given
by $|a_{Na-Rb}| \leq a_{c} = \gamma \sqrt{a_{Na}a_{Rb}}$, where
$\gamma=\sqrt{M_{Na}M_{Rb}}/(M_{Na}+M_{Rb})$
\cite{Esry1997Bohn-TBEC}. For $a_{Na-Rb}\leq -a_{c}$ the
attraction between the condensates overwhelms the repulsive
interaction within each condensate and they collapse, while for
$a_{Na-Rb}\geq a_{c}$ the mutual repulsion of the two condensates
is too great for them to overlap at all.

The single-species scattering lengths are positive for both
$^{23}$Na-$^{23}$Na and $^{87}$Rb-$^{87}$Rb collisions, which
allows for single-species BEC in either atomic species.  Using Eq.
(\ref{eq:elasticSL}) and the scattering lengths given in
Tables~\ref{tab:SL} and \ref{tab:Na&RbSL}, we calculate $a_{c}$
for all asymptotic states in a $^{23}$Na-$^{87}$Rb mixture. The
near equality of $a_s$ and $a_t$ implies that $\sigma_{el}$ will
be approximately the same for all states, and that $\sigma_{ex}$
will be small. Therefore interspecies elastic collisions will
dominate inelastic spin-exchange collisions. Using the mean values
for the triplet and singlet scattering lengths given in Table
~\ref{tab:SL}, we find that for all asymptotic two-atom states, $a
> a_{c}$.  Taking into account the known uncertainties in the various
scattering lengths, we find that the inequality $a
> a_{c}$ still holds.  We therefore conclude that
a stable, miscible TBEC in a $^{23}$Na-$^{87}$Rb mixture is not
possible.

Next, we consider the mixture $^{23}$Na-$^{85}$Rb.  This is an
interesting case, since BEC has only been observed in $^{85}$Rb by
utilizing a Feshbach resonance to tune the scattering length of
the $|F=2,m_F=-2\rangle$ state \cite{Cornish2000Wieman-85RbBEC}.
To simplify our analysis, we eliminate from consideration those
states for which the $^{85}$Rb single-species scattering length is
negative. Of the remaining states, we choose states that are
lossless with respect to both homo- and heteronuclear
spin-exchange collisions. Because of the large positive singlet
scattering length in $^{85}$Rb, there is a large variation in
$a_{el}$ from state to state.  If we use the mean values for the
interspecies scattering lengths given in Table~\ref{tab:SL}, we
again find no asymptotic states that satisfy the condition for
TBEC stability.

\section{Conclusions}

We have derived hybrid potentials for the $\Singlet$ and
$\Triplet$ states of the NaRb molecule.  We compare them to other
recently derived potentials, and we discuss why our potentials are
preferred.  We have calculated the singlet and triplet scattering
lengths from these potentials for both $^{23}$Na$^{85}$Rb and
$^{23}$Na$^{87}$Rb.  Using the elastic approximation, we have
calculated the scattering length for all two-atom asymptotic
hyperfine states for both isotopomers.  The cross sections for
elastic and inelastic spin-exchange collisions can be found using
these values.  Applying the Thomas-Fermi approximation criterion
for TBEC stability, we find no two-atom asymptotic states for
which a NaRb TBEC is stable.  Further experimental studies of
ultracold Na-Rb vapors, including efforts to produce a TBEC in the
NaRb system, will help refine our knowledge of the interatomic
potentials and test these conclusions.

\begin{acknowledgments}
This work was supported in part by the U.S. Office of Naval
Research, by the National Science Foundation, and by the U.S. Army
Research Office.
\end{acknowledgments}

\bibliography{NaRbNew_revision}

\begin{thebibliography}{38}
\bibitem[{SLf()}]{SLfootnote1}
\bibinfo{note}{The asymptotic wave function approaches a straight line for
  large $r$ if the potential falls off faster than $r^{-3}$. See Ref
  \cite{LandauQM}}.
\end{thebibliography}

\end{document}